\documentstyle[psfig,epsfig,12 pt]{article}

\def\k{\kappa}
 
\def\a{\alpha}
\def\b{\beta}
\def\d{\delta}

\def\k{\kappa}
 \def\L{\Lambda}
\def\m{\mu}
\def\n{\nu}

\def\mn{{\mu\nu}}
\def\be{\begin{equation}}
\def\ee{\end{equation}}

\begin{document}

\begin{flushright} BRX TH-500
\end{flushright}

\begin{center} {\Large\bf A Century of Gravity:
1901--2000 (plus some 2001)}

S. Deser\footnote{Invited Lecture at {\it 2001: A Spacetime
Odyssey}, Ann Arbor, May 2001} \\
Department of Physics, Brandeis University\\ Waltham, MA 02454,
USA
\end{center}

This lecture consists of two parts.  The first is a (totally
unsystematic) survey of some of the high points in the evolution
of gravity and its successors, primarily in the course of the past
century.  The second summarizes some new work on surprising
properties of higher $(> 1)$ spin fields in cosmological
backgrounds: the presence of $\L$ gives rise to discrete sets of  
massive models endowed
with gauge invariances, that divide the ($m^2 , \L$) plane into
unitary and non-unitary phases. The unitary region common to
fermions and bosons shrinks to flat space ($ \L \rightarrow 0 $)
as their spins increase.

\section{Introduction}

The scope implicit in this Conference's title seemed to mandate a
suitably sweeping topic. My initial choice, ``100$^2$, the
Sequel", would permit daring extrapolations from the present
without risk of contradiction. I quickly realized, however, that
it would also require truly long-term vision; the other
permutation, 2100, being too close, I will instead go modestly
backwards, and try to retrodict some of the high points,
especially of the past century, in the development of
gravitational physics. Indeed, this period has seen an amazing
progression, bringing the subject from a prolonged stasis to the
center (for good or ill) of almost all fundamental research since
the standard model, leaping (at least initially) over the desert
between GeV and Planck scales. Most of you will know most of this
history, and will have your own emphasis and judgements;
nevertheless it may be useful to think back to some of our smart
(if ignorant) predecessors.  In order to keep the historical
account as neutral as possible (and partly from laziness), I will
provide no references and mention no living names (you know who
you are!).

Not being a historian, however, I will balance the talk with the
usual stuff of lectures -- new, 2001, physics -- even if it means
a considerable descent from the heroic deeds of yore.  So the
second, orthogonal, part of my talk will give a brief (referenced)
account of current work, in collaboration with Andrew Waldron, on
the gravity/particle theory interface, namely the properties of
``partially massless" (the quotes are deserved) higher spin fields
in cosmological backgrounds -- deSitter (dS) or anti-deSitter
(AdS).  The new results here are quite surprising when compared
with our usual Minkowski background concepts:  There is partial
masslessness and gauge invariance at tuned values of $m^2 \neq 0$,
a phase plane with coordinates $(m^2 , \L )$ divided by these
gauge lines into unitarily allowed and forbidden regions. The
region of common boson/fermion unitarity puts bounds on, or even
forbids a cosmological constant! In presenting this research, I
also hope to satisfy those who are history-averse: they can skip
directly to it.

\section{On Whose Shoulders?}

As I launch into the past, I reemphasize that I have not tried to
be historically accurate, instead attempting to remember what
seems most seminal as one looks back over the past three centuries
(in almost as few pages!). For this reason, I begin with a
particularly impressionistic sketch of the prehistory, {\it i.e.},
the origins of modern gravitational theory.

The immediate answer to all questions is of course ``Newton", and
so it should be.  But we must give their due to the ancient
observers, particularly the Arabs, Chinese and Greeks, the
mechanicians of the Middle Ages and the geometers/arithmeticians
who helped bring mathematics to a stage sufficiently useful to
physics. It was fortunate that our solar system sufficed as an
arena, since it was both easily and secularly observable and
received enough of society's support (not always for disinterested
reasons) to provide enough data. The great predecessors nearest to
Newton (both as observers and theoreticians) include Brahe,
Copernicus, Descartes, Galileo (who also performed notable
terrestial experiments) and Kepler. I would add amongst the
theoreticians, R{\o}mer, that ancestor of special relativity, who
put $c$ on the map, unwittingly undermining Newtonian gravity at
its birth and adding his third to our triad of fundamental units.
In Newton's own shadow we must (at least) include Halley, Hooke
and Leibnitz. We should also remember the vast development,
stimulated by the Newtonian scheme, of mechanics and mathematics
by Bessel, Euler, Laplace, Lagrange, Poisson and indeed most of
the greats of the 18th and early 19th century. Their elaboration
of its ramifications, such as perturbation theory, and the
many-body problem were to be invaluable (as we have all learned)
throughout physics, even unto the birth of quantum mechanics. This
fruitful interplay between mathematics and physics (so marvelled
at by Wigner) has always been present: The Gauss-Riemann
constellation is its mid-19th century culmination, presaging (and
making possible) the geometrization of physics in a very direct
way. Hilbert, Poincar\'{e}, Weyl and their living successors have
continued the tradition. But enough ancient history; I'm supposed
to cover the 20th century!

\section{The Heroic Era:  1901--1950}

Let's, rather arbitrarily, separate the beginnings of the new
gravity from its modern and postmodern avatars.  Most conveniently
for our chronology, the obvious symbol for this division is the
Planck length, conceived, providentially, at the start of my
survey. [Planck's Note, in the {\it Berliner Berichte}, introduced
it almost as an aside.] The next few years belonged of course to
special relativity, when $c$ became a conversion factor, spacetime
as $D$=3+1 was born, but geometry remained a rigid, {\it a priori}
fixed, background.  Incidentally, it is here the seeds of $D> 4$
were sown: $D$=5 was born in flat space in 1913 when Nordstr\"{o}m
did just the opposite of what Kaluza--Klein were to propose: he
made electrodynamics part of a 5-vector in order to unify it with
scalar gravity! While it was clear after 1905 that the
instantaneous action-at-a-distance picture was doomed, that (at
the very least) $\nabla^2$ had to be replaced by
$^{\framebox[3mm]{~}}$, what emerged initially were other simple
scalar gravity models (Abraham, Mie, Poincar\'{e}) also doomed
at birth, but at least not by R{\o}mer. Only Einstein kept his
thoughts on the universality and accuracy (E\"{o}tv\"{o}s) of the
inertial/gravitational mass ratio, and very likely on the fact
that systems with vanishing $T^\a_\a$ would not gravitate in
scalar theories.

The overwhelming triumph of this period (after its gestation in
1909--1914) was of course Einstein's general relativity (GR) in
1915.  So overwhelming is this sudden revelation that we tend to
forget some of its miraculous triumphs, {\it e.g.}, its automatic
explanation of the above mass ratios -- an enormous quantitative
feat (there are a lot of very accurately measured elevators out
there!), or that because special relativity is built-in as the
local tangent space, this (mostly) guaranteed causal signal
propagation and the observed local properties of normal
non-gravitational physics.  Let us also pay tribute to that
incredibly useful invention, the summation convention (whoever has
seen Einstein's notes from before and after this illumination can
vouch for its effect on his work, let alone ours!).  Einstein's
introduction, already in 1917, of the cosmological constant $\L$
showed his instinctive understanding that, in physics, all that is
not forbidden is compulsory (or at least we don't yet have a
natural forbiddenness for $\L$, except in one model, unbroken
$D$=11 supergravity).

Geometry$\Leftrightarrow$Dynamics has been {\it the} pillar of all
subsequent unification dreams.  Indeed, as in all matters,
Einstein was ahead of his time: after writing $G_\mn = \k^2
T_\mn$, he immediately asked who ordered the right-hand side.
Unfortunately for him, it was the right question at the wrong time
and was to lose him in his later years, but we all still pursue
it, with (we hope) deeper wisdom: some version of it will always
be {\it the} question. Let me cite a much less known example of
Einstein's foresight to show that great steps also engender
subtleties: Very early on he realized the dangers inherent in a
dynamically determined geometry: how, he asked, could we be sure
that it would always be well-behaved, namely causal?  He
understood the horrors of time travel and predicted that ``good"
sources would somehow not generate ``bad" spaces, and so it has
turned out.

The next triumphs were Schwarzschild's rigorous black hole
solution the following year, showing also that global properties
count (as was realized only long after), actually preceded by
Einstein's description and calculation of the famous three tests.
While we usually say that of these, only perihelion precession
probes post-linearized effects, that is a bit narrow in that
linearized theory is a fundamentally inconsistent (if very useful)
model. [Birkhoff's theorem, in this connection, was an early way
to say there would be no monopole or dipole radiation.]

Consistent cosmology started in the early twenties, with Friedmann
and Lemaitre as some of its pioneers. In the same period, came the
next big generalization: geometry-driven Kaluza--Klein unification
by use of higher dimensions, for them the simplest step,
$D\rightarrow 5$. After suffering many swings of fashion, $D
> 4$ (or even $D >> 4$) seems to be here to stay, as are indeed
both $D$=3 and $D$=2! [As everyone at this Conference knows, the
more sophisticated, Klein, version was born in Ann Arbor. Klein
was simultaneously teaching GR and electromagnetism and tried to
lighten his load by this stratagem; great as that version of
history would be, the truth is that he was actually inspired by
teaching mechanics, in fact the Hamilton--Jacobi method in
background gravitational and electromagnetic fields.] Note also
the concomitant introduction of topology, {\it i.e.},
compactification of the fifth dimension, and of global geometry to
quantize a physical parameter, here electric charge, a precursor
to the quantization of other constants, such as the non-Abelian
Chern--Simons coefficients.

After this intensively successful period, progress slowed, partly
because of the irresistible appeal of quantum mechanics, QFT and
particle physics, partly because the broad lines of the classical
theory had been understood and no earthshakingly novel
observations were at hand (except, as usual, in cosmology: Hubble
expanded our world already in 1929). Indeed, things got moving
again in a big way when GR begun to be investigated, first as a
classical, and then more painfully as a quantum field theory, in
the late fifties. To be sure, there was plenty of important prewar
progress. A short list would include: the expression, by Fock,
Klein, Schr\"{o}dinger, Weyl and others around 1929 of spinor
theory in curved space, a formal but significant start towards
quantum gravity, let alone supergravity! Then there was (in no
particular time order) the work of Oppenheimer and his school, and
Chandrasekhar's, on black holes, while ``time travel" solutions,
most notably G\"{o}del's provided suitable puzzles. On the
mathematical side we may cite the beginning of rigorous analysis
of the Einstein equations as a normal hyperbolic system, with
propagating and perfectly real gravitational waves (something that
was not at all obvious to many people at the time) by Darmois,
Lichnerowicz and his school amongst others. Perhaps more
significant to the quantum side of our story were the early
recognitions, by Rosenfeld already in 1930, that GR had to be
quantized if only because its matter sources are, and by
Heisenberg in 1938 that the dimensionality of the Einstein
coupling constant boded ill for the ultraviolet behavior of
quantum gravity, just as it did for the then new Fermi weak
interaction model.  The thirties also saw the rebirth of the
linearized approximation as the field theory of massless spin 2 at
the hands of Fierz, Pauli and independently by Bronshtein in the
USSR (before he disappeared in the Stalin purges).  We must also
not forget the modern understanding of the stress-tensor as a
(covariantly) conserved current by Belinfante and Rosenfeld.

Important as it was, the progress I have traced from say 1925 to
WWII was distinctly low-budget, with few adepts: it was a very
unfashionable path to follow, and suffered from the absence of new
experimental input as compared to lower spin physics. [To be sure,
while Einstein didn't discover GR due to immediate pressure of new
data, his incorporation of hitherto unexplained observations
certainly qualified as being experimentally driven.]  What were
the reactions of the great quantum innovators to GR? Some of the
giants of the quantum era later turned into relativists --
Schr\"{o}dinger was perhaps the first (in fact he had also worked
in GR long before 1925) although he foundered on the same dead end
as Einstein; Dirac started much later and stayed (mostly) within
the existing theory as a dynamical system, but was also driven by
the ``large number problems" to consider gravity with variable
gravitational constant. Pauli, whose breakthrough was (at 18) with
one of the first texts on GR, essentially stayed on the sidelines
thereafter. Jordan is best remembered for the scalar-tensor theory
and Klein remained active in both $D$=4 and 5.  Others, like Bohr,
Born, Heisenberg, Wigner and Planck himself, never really entered
GR. Fermi's very first papers (he was inspired as a young student
by the Italian geometry school) were in GR, and as might be
expected, are useful to this very day. The converse reaction was
Einstein's refusal to enter the promised land of QM!

\section{Modern and Postmodern 1950--2000}

With the general postwar physics upsurge came a new wave here as
well. Initially, much talent was deployed in generating, and
classifying types of solutions; this has remained an industry,
with such beautiful results as the rotating black hole geometry
(that seems just now to have found its observational
realization!), chaotic cosmologies, ``radiative" metrics and their
geometric representations, as well as the deepening understanding
of black holes. From the late fifties, GR began to be studied as a
gauge field theory, to which physical concepts such as energy and
the usual apparatus of Hamiltonian dynamics could be applied,
under appropriate asymptotic behavior. Likewise, the covariant
formulation of the theory, including the (index-heavy)
perturbative calculational rules began to be used for tree, and
later loop, calculations. Indeed the complications peculiar to
nonabelian theories, such as ghosts and their role in keeping the
rules free of unitarity paradoxes were understood here in the
early sixties. The seventies brought in a wider integration into
the rest of QFT. For example, gravitational conformal anomalies
began to be explored by the mid-seventies.  These anomalies were
in turn connected to black hole radiation as part of a new (and
still very strong) industry of QFT on a fixed gravitational
background, that enabled progress to be made in quantized
matter-gravity interactions.

There was also a dark side to the new results: as Heisenberg had
feared, the dimensional coupling constant of GR indeed gives rise
to catastrophic ultraviolet loop effects.  [The flip side is that
the infrared end is totally unproblematic, precisely because
stress tensors, {\it i.e.}, vertices, contain a higher power of
momentum than vector currents do.]  Let's recall rapidly what the
generic problem is, say for a bosonic field (fermions are
similar):  The vertex (stress-tensor) behaves as $\sim p^\m p^\n$,
while the propagator is $\sim p^{-2}$.  Thus the insertion of an
extra internal line into a given loop order diagram involves two
new vertices and three new denominators, together with a new
virtual $d^4k$ integration, a process that (barring miracles)
leads to increasingly virulent infinities and hence counterterms
with increasing powers of curvature at each loop order.  That this
is also the case in practice was rapidly borne out in three
different sectors.  First, the calculational rules were
sufficiently streamlined to show (almost thirty years ago) that at
one loop level gravity plus a scalar field is already
non-renormalizable, requiring non-vanishing new counterterms.
While pure GR, itself a self-interacting quantum system, also led
to new infinities, by an accident of $D$=4 geometry, the
corresponding terms could be removed by field redefinitions since
they involved local terms quadratic in curvature.  All these terms
vanish when the external gravitational lines are on-shell $(R_\mn
= 0)$ by the Gauss--Bonnet identity, which reduces the three
possible quadratic terms (including $\int R^2_{\mn\a\b})$ to just
$\int R^2$ or $\int R^2_\mn$.  This ought no longer to be the case
at 2 loops, but here the calculations become almost insuperable.
Nevertheless, they were carried out around 1986 and the
corresponding unremovable $R^3_{\mn\a\b}$ counterterms were
present, as was independently verified a few years later.  The
final nails in the gravity plus matter coffin were driven in by a
series of one-loop calculation (also in the early seventies)
including Dirac particles, photons and Yang--Mills fields coupled
to gravity: even for matter gauge theories there were no miracles
at all, bearing out the need for something new, at least within
the perturbative framework that is the only available tool in QFT.
The inescapable conclusion is that GR (like the Fermi theory) is
at best an effective low energy theory, and must be the limit of a
more fundamental one.

Sure enough, something, indeed two new things, did show up.  The
first, string theory, had come up a bit earlier, emerging from
strong interaction theory (the dual resonance model). That closed
strings contained massless spin 2 excitations in the zero slope
limit and that these could, indeed must, be gravitons was duly
noted but remained unexploited until the string revolutions began
in the '80s. Less exotic, but more immediately accessible, was of
course Supergravity (SUGRA), the local gauge generalization of
ordinary flat space supersymmetry's invariance under constant
spinor transformations, turning the graviton into part of a
doublet with a spin 3/2 field. Found exactly 25 years ago this
Spring, the so called $N$=1, $D$=4 model was rapidly generalized
both in dimension and in field content. At first it seemed just
the ticket to finiteness: not only did it share pure GR's 1-loop
escape, but unlike GR it did not permit 2-loop
(super-)gauge-invariant counterterms.  Alas, it was shown very
soon to have plenty of possible ones at 3 loops and up! So in this
sense SUGRA was no better a gauge theory than GR plus ordinary
matter, except to defer the infinities to a higher loop order.
Indeed, quite recently it was shown explicitly that even the most
beautiful and ultimate SUGRA, that of $D$=11, is already 2 loop
nonrenormalizable, requiring specifiable infinite counterterms.
Thus, the short-term dream for SUGRA as both a unified and finite
theory has proven invalid at least as to finiteness. Nevertheless,
many people continue to share the hope that $D$=11 SUGRA, so
unique that it cannot have matter sources at all (a beautiful
realization of Einstein's dream about no right-hand side of the
ultimate field equations) is, in some broken version, at least
part of the next big step.

It is impractical to attempt any detailed account of gravity's
progress (or at least annexation!) beyond this period: the
postmodern era is very much unfinished business on the string,
M-theory, brane and novel compactification fronts.  We have gone
long beyond pure GR, but with no realistic unification in place
yet. This period has however been full of observational (and
associated theoretical) triumphs. Three examples are: (1) the
binary pulsars and gravitational radiation, with total and
extremely accurate agreement with GR calculations. (2) The
development and increasingly accurate vindication of the big bang
picture, albeit with many novel puzzles from our apparently living
in a deSitter cosmology to the fact that known matter is
practically ``in the noise" compared to as yet unknown but
seemingly necessary new forms of it. (3) Some form of inflation as
the key to ``homogeneous and isotropic". We await the start of
gravitational wave observations from LIGO and its several
acronymic counterparts into a new form of astronomy, and expect
continued improvement in a variety of observational regimes, both
here and aloft.  All this information will bring in a new
phenomenology, if not yet the desired superunifications.  So the
new century (let alone millennium), can be safely predicted to be
an exciting one!

\section{Progress and Mysteries; a Summary}

Let me close this survey with a (random) summary of what was said
about the current big picture and of its mysteries. Some main
lessons from the evolution of gravity have been:
 \begin{enumerate}
 \item
 Instantaneous scalar, linear, Newtonian gravity has been
naturally required by special relativity and universality of
coupling (including notably light bending) to transform into a
propagating tensor theory, and (on consistency grounds)
necessarily a self-interacting one. This necessity immediately
yields Einstein theory, {\it i.e.}, geometry, as was shown
explicitly some 3 decades ago: dynamics and geometry are
unavoidably unified.
 \item
 Cosmology acquires a natural basis; and, for better or worse, a
cosmological constant can be present. Many observed phenomena in
the large find a satisfactory underlying explanation, others just
the opposite.
 \item
 Geometry as the generic road to unification. Invariances specify
 the correct dynamics in all areas, including the
 non-Abelian gauge theories of the standard model: there is now just one (generalized)
 T-shirt motto, $[D_\m ,  D_\n ] = F_\mn$.
 \item
 Gravity reemerges as an experimentally driven science: the 3
 ``old" tests, supplemented by two more recent ones, time delay and lunar
 ranging; binary pulsars and the reality of gravitational waves;
 black hole observations, and of course the mapping of the big
 bang cosmos are just some examples.
 \item
 The importance of thinking globally: topological invariants,
 coordinate identifications, and understanding of global
 properties of {\it e.g.}, black holes; singularity theorems;
 Lorentz generators in asymptotically flat spaces, $D$=3 gravity
 where dynamics is just global geometry, Chern--Simons terms in
 (odd-dimensional) physics.  Finally, we mention $D$=2 (super)gravity
 where only ``the smile" is left, but just enough to generate the (super)
 string  action in a precise way through its (lower spin) matter sources.
 \item
 The liberation of dimensions: $D = 4 + n$ as a post-Einstein road
 to geometrical unification, with mere Kaluza--Klein $n$=1 now
 replaced by (at least) $n$=6 or 7.  The sensitivity of
 supergravity to $D$, with its natural upper limit $D$=11, along
 with superstrings' $D$=10.  Compactifications on wildly
 different (not just KK Planck) scales, and with new building blocks, branes.
 \item
 Gravity and Geometry redux: GR and SUGRA as mere local limits of
 much broader systems, {\it e.g.}, strings or M-theory or their
 non-commutating generalizations.
 \end{enumerate}
 The above list is both incomplete and not at all meant to convey
relative importance in terms of prospects for the future survival
or incorporation into the next breakthroughs, of any or all of these ideas.

With all really deep progress comes a set of mysteries and gravity
is no exception.  Here is one such enumeration:
 \begin{enumerate}
 \item
 One of the true formal triumphs that deserved to have its own
 place in our first list is certainly SUGRA, particularly its role
 as the ``Dirac square root" of geometry.  Just knowing that GR
 has such a square root has given us enormous insights, for
 example that its energy must be positive and flat space its
 vacuum.  Yet after a quarter-century
  SUGRA's role and scope have not yet been fully
 understood, particularly its most mysterious maximal $D$=11
 version.  That $D$ must be bounded by 11 is really a consequence of the
 well-known problems (unavoidable in $D>11$) of coupling higher
 $(s>2)$ spin gauge fields to gravity.  Rather, the mysteries seem
 to lie in the $D$=11 theory's uniqueness:  neither
 supersymmetric ``matter" (lower spin multiplets) to provide a
``right-hand side", nor even a cosmological term can be present.
It requires a (cubic) Chern--Simons term for its 3-form field, one
that is (perversely) $P$-even.  Is this model really a ``mere" QFT
(remember that all SUGRAs are necessarily quantum rather than
classical theories due to the fermionic $s=3/2$ companion of the
graviton) or (as is now thought) on a footing with {\it e.g.},
string theories? We do know, as mentioned earlier, that it is {\it
not} a magically perturbatively finite QFT.
\item
  Compactification:  the freedom of living in $D>4$ both in QFT
  and in string theory brings with it the onus of providing a
  credible mechanism for landing in the observed corner that is
  our $D$=4 world.  This has been one of the leading challenges of
  the past decades.
  \item
  Black hole physics.  Despite tantalizing clues, and much
beautiful speculation, it seems fair to say that we do not yet
have sufficient understanding of the statistical mechanics basis
for thermodynamic laws, nor of the black hole radiation and
  information loss problem, or of the holography -- surface degrees of
  freedom connection, amongst others.
  \item
  The cosmological constant problem.  This one is too well-known
  to require mention, but apart from the grotesque discrepancy
  between naive zero-point matter energy and the observed size of
the universe, things are in some ways worse now that observation
  seemingly dictates a non-vanishing deSitter (dS) value of $\L$.
  This means that any future fundamental way to set $\L$ strictly
  to zero will now require a plausible symmetry breaking.  In
  addition, if we must accept $\L > 0$, we must also cope with the formal
  horrors of doing physics
  in a dS universe with its intrinsic horizon and other
  ``non-Minkowskian" properties (not that $\L < 0$, anti-dS, is
  any better!).
\item
 Assuming that string, or M-theory or their successors do
 indeed incorporate classical GR in a satisfactory and finite way,
 in that it indeed emerges as some zero slope effective limit,
 then we are still stuck at the old desert crossing
 dilemma: Do these deeper ``geometries" really also contain the
 physics of the standard model in a unique, recognizable way, and
 if so how is the gap of 18 or so orders of magnitude accounted
 for?  It is after all quite possible that we have been
 deluding ourselves in assuming that whatever legitimizes GR (or
 SUGRA) by incorporating it in some consistent quantum scheme
 necessarily also includes precisely the matter we study at the standard
 model level.  This criticism, often heard in the early days of
 string theory, is not yet entirely obsolete.
 \end{enumerate}

The above problems, and especially those we have not even
conceived of, should provide fun for the next century.  Of course,
the {\it real} fun will be the rest of our millenium when we try
to incorporate (or do away with) the other -- sub-Planck scale --
desert!  But as Bohr reminded us, prediction, especially of the
future, is treacherous  and he even proved it: a couple of years
before the 1925 QM revolution, Bohr was supposed to have doubted
that any true improvement over the old quantum theory would come
in his lifetime.  Might the (next) TOF also be just around the
corner?

 \section{A 2001 Sample:  Surprises from Higher Spin Fields
 in (A)dS}

There should be at least some original physics even in a survey,
and as threatened, here is my 2001 contribution. I propose to
summarize some current work \cite{001,002,003,004} at the
gravity-particle interface, namely exotic properties of ordinary
familiar free fields living in cosmological spaces.

Owing to (another kind of) space constraints, I will begin at the
end, namely state some of the main results: A non-vanishing $\L$
(for us $\L$ has dimension $m^2$) necessarily alters the
kinematics of free massive higher $(s>1)$ spin fields in such a
way that (for appropriate discrete values of the $m^2 /\L$ ratio)
there are ``partially massless" models, whose novel gauge
invariances can successively eliminate all possible sets of lower
helicities: These eliminated sets not only include all those from
$(s-1)$ down to zero or 1/2 as in the usual flat space massless
gauge models, but also just helicity 0 or 1/2, or (1/2,3/2) or
(0,1) all the way up. These discrete ``tuned" lines act as ``phase
transition"  in the $(m^2 , \L )$ plane, separating regions that
are either forbidden and permitted according to whether they
contain excitations that violate unitarity or not: Essentially,
there is only one allowed, unitary, region for bosons and one for
fermions. The two are mostly complementary, but possess an
interesting overlap about the $\L = 0$ axis in this plane.  In the
limit, as fields of arbitrarily high spin (independent of details
of the masses) are included, the unitarity region on each side
narrows precisely to the $\L = 0$ line:  no $\L \neq 0$
cosmological constant space can permit infinite higher spin
towers.  Here is the picture that is worth all the above words:

\vspace{1.cm}
\begin{center}
\epsfig{file=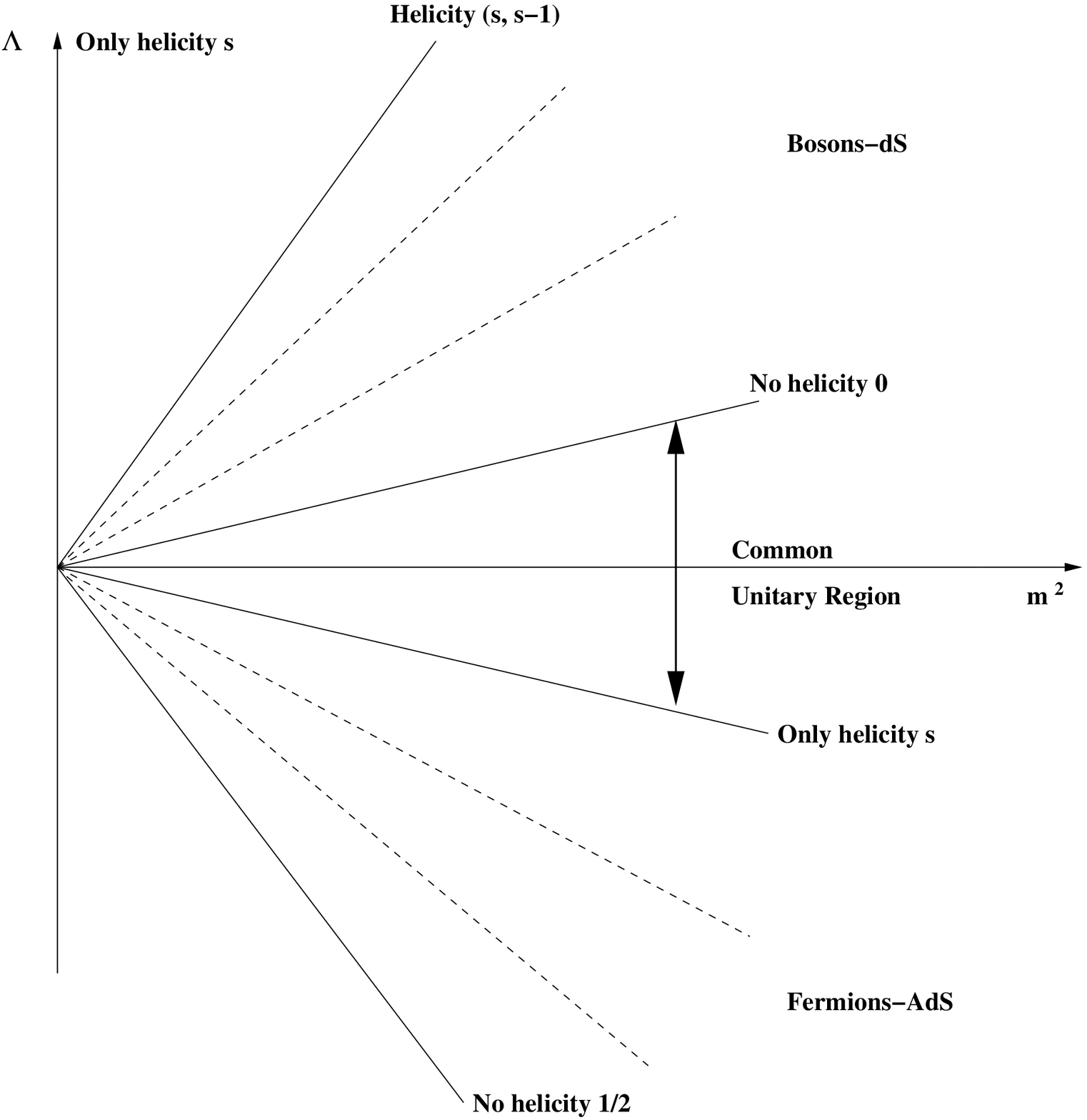, width= 15.cm}
\end{center}
\baselineskip=13pt \centerline{}

\noindent\footnotesize{\bf Figure 1}:  The top/bottom halves of
the half-plane represent dS/AdS (and also bosons/fermions)
respectively. The $m^2=0$ vertical is the familiar massless
helicity $\pm s$ system, while the other lines in dS represent
truncated (bosonic) multiplets of partial gauge invariance: the
lowest has no helicity zero, the next no helicities $(0,\pm 1)$,
etc.  Apart from these discrete lines, bosonic unitarity is
preserved only in the region below the lowest line, namely that
including flat space (the horizontal) and all of AdS. In the AdS
sector, it is the topmost line that represents the pure gauge
helicity $\pm s$ fermion, while the whole region below it,
including the partially massless lines, is non-unitary. Thus, for
fermions, only the region above the top line, including the flat
space horizontal and all of dS, is allowed.  Hence the overlap
between permitted regions straddles the $\L = 0$ horizontal and
shrinks down to it as the spins in the tower of spinning particles
grow; only $\L = 0$ is allowed for generic ($m^2$ not growing as
$s^2$) infinite towers.

 \normalsize Specifically, one may deduce
\cite{004} from the fact that the transition line gauge models can
be shown to all have null propagation (another remarkable property
of these partially massive theories) that the unitary region is
bounded by the lines
 \be
 3m^2_b = \L \: s (s-1) \; ; \;\;\; 3m^2_f = - \L
 (s-1/2)^2
 \ee
respectively. [Since the relevant bosons/fermion regions are
dS/AdS $(\L > 0/\L < 0)$, the signs in (1) are correct as shown.]
Clearly, as $s$ increases, then (unless the masses conspire to
rise as high as quadratically with spin) the allowed $\L$  must
clearly vanish in both half angles above, {\it i.e.}, infinite
towers of higher spins can only exist in $\L = 0$ spaces. Whether
the above mechanism has any relevance to the
 infinite towers of higher spins in string slope expansions is far
 from clear.  It is also not clear how robust it is within a QFT
 framework, once interactions and dynamical gravity are turned on.
But this dramatic picture certainly provides a sample of how
particle kinematics can be affected by cosmological backgrounds,
allowing only certain partial gauge theories and a restricted
range of $m^2$ for a given $\L$.

How does all this ``level-splitting" compared to the $\L = 0$
picture arise?  At the action level (rather than in terms of
representations of the appropriate (A)dS algebras), it is due to
lifting of degeneracy among successive Bianchi identities on
higher spin field operators.  I can only sketch the simplest
example here, spin 2, the lowest nontrivial case. In a dS
background (denoted by a bar), the successive divergences of the
field equations read
$$
\bar{D}_\m \, {\cal G}^\mn = -m^2 \, \bar{D}_\m \, (h^\mn -
\bar{g}^\mn h^\a_\a )\; , \eqno{(2\rm{a})}
$$
$$
\bar{D}_\m \bar{D}_\n \, {\cal G}^\mn + \frac {1}{2} \, m^2 {\cal
G}^\m_\m =  3/2 m^2 \, (m^2 - 2/3 \, \L ) h^\a_\a \; .
\eqno{(2\rm{b})}
$$
where $\cal{G}^\mn$ is the linearized Einstein tensor  of the
massive field $h_\mn$ (it is unique if we require $\bar{D}^\m
{\cal G}_\mn \equiv 0$). The first divergence is as in flat space
and eliminates helicity $\pm$1 if $m^2 = 0$. However, the second
divergence shows that while helicity 0 is eliminated as well at
$m^2 = 0$ (the usual massless spin 2 gauge invariances that leave
only helicity $\pm$2), it is also possible \cite{005} to tune
$m^2$ to $2/3 \, \L$, eliminating helicity 0 but {\it not}
helicity $\pm$1, and leaving only the 4 excitations $(\pm 2, \:
\pm 1)$, due to a novel (scalar) gauge invariance at this point,
namely under
\renewcommand{\theequation}{\arabic{equation}}
\setcounter{equation}{2}
 \be
\d h_\mn = (\bar{D}_\m \bar{D}_\n + \bar{D}_\n \bar{D}_\m + 2\L /
3 \bar{g}_\mn ) \a  \; .
 \ee
That the region lying between the $m^2 = 0$ and $m^2 = 2/3 \: \L$
line $s$ is non-unitary was discovered in \cite{006} by
considering equal time commutators among field components and
their behavior as a function of $(m^2 , \L )$. It can be
understood in a different way by tracing the behavior of the
helicity 0 mode in the $(m^2, \L )$ phase of Figure 1. Clearly,
the theory has five normal flat space helicity excitations (for
$m^2 \neq 0$) at $\L = 0$, the horizontal line. As one comes up in
dS space, the helicity 0 mode first vanishes on $m^2 = 2/3 \, \L$
and then reappears on the other side, but with ghost behavior (see
below). This loss of unitarity affects the whole region until we
reach the $m^2 = 0$ line, where the helicity 0 mode again vanishes
(along with helicity $\pm$1);  this limit is just linearized dS
gravity, which was analyzed long ago \cite{007}.

Let me sketch finally a simple but I trust instructive Hamiltonian
analysis \cite{003} of how this occurs:  The free theory's action
is (as usual) the sum of those of the separate helicities. Because
helicity $(\pm 2, \pm 1)$ modes are entirely independent of the
linear, scalar, Hamiltonian constraint, only the helicity zero
part need be considered. After some field redefinitions (canonical
transformations) the helicity zero Lagrangian reduces to the form
 \be
 L_{h=0} (p,q) \equiv p\dot{q} - \textstyle{\frac{1}{2}} \, [\n^2
 p^2 + \n^{-2} q (-\nabla^2 + \m^2 )q ]
 \ee
 where $\n^2 \equiv m^2 - 2\L /3$, $\m^2 \equiv m^2 -
 \textstyle{\frac{3}{4}} \L $ and $\nabla^2$ contains a
 time-dependent factor $f^{-2}(t)$ in the synchronous gauge we
 are using here,
 \be
 ds^2 = - dt^2 + f^2 (t) dx_i^2 \; , \;\;\; \ln f = \sqrt{\L /3} \:
 t \; .
 \ee
The shift in mass value from $m^2$ to $\m^2$ is essential both to
establish (non-trivially) positive energy and null propagation
\cite{004} at both lines, but in the present context, $\n^2$ is
the operative factor.  So long as $\n^2 > 0$, we may trivially
remove it by an obvious rescaling of $(p , q)$ and obtain a
perfectly unitary model.  But when $\n^2$ becomes negative (the
``bad" region) then we can no longer rescale $(p,q)$ by $\n$
without either introducing imaginary fields or rescaling by $|\n
|$ and accepting a negative helicity zero Hamiltonian.  The
quantum version of this non-unitarity also emerges, as mentioned,
from a study of the equal time commutators \cite{006}, that
displays a concomitant ghost for $\n^2 < 0$. Precisely at $\n^2 =
0$, the whole excitation is easily shown to vanish even before the
desired form (4) is even reached, just as helicity 0 quickly drops out
of vector actions precisely at $m^2=0$ in flat space. Entirely
similar results hold for higher integer spins where more indices
allow for more sets of helicity deletions (starting with the
lowest), as well as (in AdS) for spinors with $s > 3/2$, as
symbolized in our Figure.

It is a pleasure to thank A.\ Waldron for intensive and extensive
collaboration. I am grateful to several colleagues, especially T.
Damour, F. Hehl and A. Trautman for helpful historical lore. This
work was supported by NSF Grant PHY99-73935.


\begin{thebibliography}{99}
  \bibitem{001}
  S. Deser and A. Waldron, Phys. Rev. Lett. {\bf 87}, 031601 (2001).
\bibitem{002}
  S. Deser and A. Waldron, Nucl. Phys. {\bf B607}, 577--604 (2001).
\bibitem{003}
  S. Deser and A. Waldron, Phys. Lett. {\bf B508}, 347--353 (2001).
\bibitem{004}
  S. Deser and A. Waldron, Phys. Lett. {\bf B513}, 137--141 (2001).
\bibitem{005}
 S. Deser and R. Nepomechie, Phys. Lett. {\bf B132}, 321 (1983); Ann.
 Phys. {\bf 154}, 396 (1984).
\bibitem{006}
 A. Higuchi, Nucl. Phys. {\bf B282}, 397 (1987); {\it ibid.} {\bf
 325}, 745 (1989); J. Math. Phys. {\bf 28}, 1553 (1987).
\bibitem{007}
 L.F. Abbott and S. Deser, Nucl. Phys. {\bf B195}, 76 (1982).
   \end{thebibliography}
\end{document}